\documentclass[groupedaddress,twocolumn]{revtex4-2}

\usepackage{graphicx}% Include figure files
\usepackage{xcolor}
\usepackage{dcolumn}% Align table columns on decimal point
\usepackage{amsmath}
\usepackage{amssymb}
\usepackage{bm}

\begin{document}

\title{Strong Mitigation of the Magnetic-Field-Induced Frequency Shift\\in Coherent-Population-Trapping Atomic Clocks}

\author{Denis Brazhnikov}
 \email{Contact author: x-kvant@mail.ru}
 %\altaffiliation[Also at ]{Physics Department, XYZ University.}%Lines break automatically or can be forced with \\
\author{Vladislav Vishnyakov}%
% \email{Contact author: Second.Author@institution.edu}
\author{Mikhail Skvortsov}
\affiliation{Institute of Laser Physics SB RAS, 15B Lavrentyev Avenue, Novosibirsk 630090, Russia
}%

\date{\today}

\begin{abstract}

We study the magnetic-field-induced frequency shift (MFS) of the clock (``0--0'') transition in coherent-population-trapping (CPT) microwave atomic clock. It is shown that the use of the Pound-Drever-Hall-like (PDH) technique for frequency locking provides brilliant opportunities for mitigating the MFS. Using a $0.125$~cm$^3$ rubidium vapor cell with a buffer gas, we have measured a residual sensitivity of the clock transition frequency to be $\approx\,72$~$\mu$Hz/mG over $\approx\,$$6$ mG interval. It means that a fractional frequency shift is extremely small ($\approx\,$$1$$\,\times\,$$10^{-14}$~mG$^{-1}$). The results contribute to the development of a new-generation CPT-based miniature atomic clock (MAC) with improved long-term frequency stability. The proposed method is quite general and can be used for other excitation schemes in atomic clocks, including Ramsey-like techniques.

\end{abstract}

%\keywords{Suggested keywords}%Use showkeys class option if keyword
                              %display desired
\maketitle

\textit{Introduction.}---Miniature atomic clocks (MACs) based on coherent population trapping (CPT) in alkali-metal vapor cells \cite{Vanier2005,Kitching2018}, owing to their small size ($V$$\,<\,$$100$~cm$^3$), weight ($M$$\,<\,$$100$~g) and power consumption ($P$$\,<\,$$1$~W), have already found important applications, such as global navigation satellite systems \cite{Meng2024}, remote sensing \cite{Aheieva2017} and others \cite{Cash2018}. MACs are designed on the basis of a microwave transition in $^{87}$Rb ($6.8$~GHz) or $^{133}$Cs ($9.2$~GHz), which is often called ``0--0'' or ``clock'' transition. This microwave transition is induced by two coherent optical frequency components in the spectrum of a vertical-cavity surface-emitting laser (VCSEL), avoiding the use of any resonant microwave radiation. This all-optical technology makes MACs highly compact and energy efficient.

The cutting edge MACs demonstrate a fractional long-term frequency stability (Allan deviation) of $\approx\,$$2$$\times$$10$$^{-12}$ at 24~h \cite{Zhang2019,Skvortsov2020}. The efforts of many research groups are nowadays aimed at increasing further the frequency stability of MACs, while maintaining their small size, weight and power consumption (SWaP). In practice, several major factors limit long-term frequency stability of MACs, which are associated with the random variations and drifts in such quantities as temperature of the atoms ($T$), optical power in the vapor cell ($P$), microwave modulation power of VCSEL's pump current ($P_\mu$) and ambient magnetic field ($B$). These variations result in mid-term and long-term drifts of the ``0-0'' transition frequency. Buffer gas permeation through the cell glass windows is also a limiting factor. However, it can be overcome by employing alumino-silicate glass (ASG) \cite{Dellis2016,Carle2023}.

The most obvious step to improve frequency stability is to stabilize all the above listed parameters to minimize their variations over time \cite{Vicarini2019}. Unfortunately, this measure is not sufficient to achieve long-term stability at the level of $10$$^{-12}$, and it is even less sufficient for further improving stability down to $\sim$$10$$^{-13}$ in next-generation MACs. In the case of a continuous-wave regime for the CPT resonance excitation, a common strategy to reduce the sensitivity of the clock transition frequency to variations in $T$, $P$, $P_\mu$ and $B$ consists in identifying specific points in the dependence of the clock transition frequency shift ($\Delta$) on each of these parameters. In particular, when using a specific mixture of buffer gases in the vapor cell, $\Delta(T)$ exhibits an extremum at a certain $T$$\,=\,$$T_0$, which determines the operating temperature of the cell \cite{Skvortsov2020,Vanier1982,Kozlova2011}. In vicinity of this extremum, the function $\Delta(T)$ has a quadratic manner, making the frequency of the atomic clock much less sensitive to small variations in vapor temperature than at any other value $T$$\,\ne\,$$T_0$. Similarly, it is possible to find such a microwave power $P_\mu$$\,=\,$$P_{\mu1}$ at which the function $\Delta(P_\mu)$ exhibits an extremum \cite{Levi2000,Skvortsov2020,Vaskovskaya2024}. At another microwave power $P_\mu$$\,=\,$$P_{\mu2}$, $\Delta(P)$ behaves as a horizontal line, indicating the absence of sensitivity of the clock transition frequency to changes in $P$ \cite{Levi2000,Miletic2012,Radnatarov2021,Vaskovskaya2024}.
There two novel schemes have been proposed with $P_{\mu1}$$\,=\,$$P_{\mu2}$ \cite{Vaskovskaya2019, Brazhnikov2024}, which gives hope that in the future, the light shift influence will be suppressed down to $\sim\,$$10^{-13}$.

\begin{figure*}[!t]
\includegraphics[width=1\linewidth]{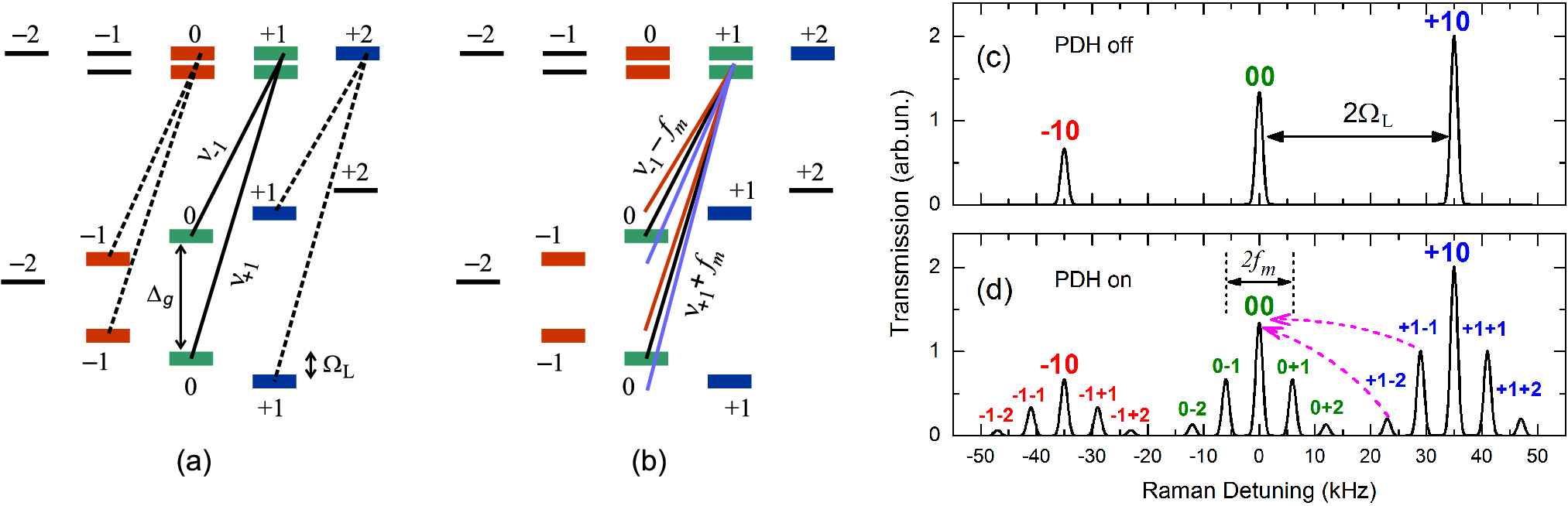}
\caption{\label{fig:1}(a), (b) Optical transitions $F$$\,=\,$$1,$$\,2$$\,\to\,$$F'$$\,=\,$$2'$ in the D$_1$ line of $^{87}$Rb ($\lambda$$\,\approx\,$$795$~nm) induced by the circularly polarized ($\sigma^+$) frequency components $\nu_{-1}$ and $\nu_{+1}$ of VCSEL radiation. A bias magnetic field in the vapor cell causes splitting of energy levels, which is shown only for the ground-state hyperfine levels. Solid lines in (a) correspond to the formation of the clock CPT resonance ``$00$'' in (c), while dashed lines are responsible for the magnetically-sensitive CPT resonances located to the left (``$-10$'') and right (``$+10$'') of the clock resonance. In (b), the PDH technique is used to excite the atoms. The laser spectrum acquires a few additional frequencies such as $\nu_{\pm1}$$\,\pm\,$$f_m$. To simplify the figure, we show only those transitions which are responsible for the central group of resonances in (d). Simulated lineshape of the CPT resonance at $f_m$$\,=\,$$0$ (c) and $f_m$$\,=\,$$6$~kHz (d). Other parameters of the simulation: $2\Omega_{\rm L}$$\,=\,$$35$~kHz, $\Gamma_i$$\,=\,$$\Gamma_{ij}$$\,=\,$$100$~Hz (here, $i$$\,=\,$$1,2,3$; $j$$\,=\,$$1,$$\dots$$,5$), $A_1$$\,=\,$$1$, $A_2$$\,=\,$$2$, $A_3$$\,=\,$$3$, while the amplitudes of the remaining resonances in (d) represent a certain proportion of the amplitudes $A_{1,2,3}$.}
\end{figure*}

To reduce the influence of magnetic field variations on frequency stability of MACs, a bias dc magnetic field (usually larger than $100$~mG) is applied along a wave vector of the beam \cite{Skvortsov2020,Knappe2007}. We assume this field to be directed long the $z$-axis ($B_z$). In such a configuration, the ``0--0'' transition in the atom can be excited separately from other microwave transitions. A key feature of this transition is that its frequency is immune to small magnetic field, retaining only a quadratic Zeeman (QZ) shift \cite{Vanier2005}. At the same time, since a relatively large bias field is applied to the atoms, the ``0--0'' transition frequency acquires a linear-like sensitivity to small perturbations in $B_z$ \cite{Knappe2007,Hong2020}. Indeed, the QZ shift reads

\begin{equation}\label{eq:QZshift}
\Delta_{\rm QZ} = \varkappa(B_z + \delta B_z)^2 \approx \varkappa B_z^2 +2\varkappa B_z \delta B_z\,,
\end{equation}

\noindent where $\varkappa$$\,\approx\,$$575$~Hz/G$^2$ for $^{87}$Rb and $\delta B_z$ is the small perturbation in $B_z$ ($\delta B_z$$\,\ll\,$$B_z$). This issue is commonly addressed in MACs through a passive magnetic shielding of the vapor cell with a factor at best of $1000$ \cite{Knappe2007}. In the case of chip-scale atomic clocks this is a challenging task. In this way, further improvement in frequency stability of MACs requires novel approaches for mitigating the magnetic-field-induced frequency shift (MFS).

In Ref. \cite{Tsygankov2021}, the authors revealed that the function $\Delta (B_z)$ exhibits a single extremum, which can serve as a bias magnetic field in MAC. Other recent experiments have shown that the use of the Pound-Drever-Hall-like (PDH) technique for frequency locking expands the possibilities for controlling the behavior of $\Delta (B_z)$ \cite{Vishnyakov2025}. According to this technique, to form an error signal, the CPT resonance is modulated at a frequency $f_m$, which significantly exceeds the resonance linewidth (FWHM is typically about $1$~kHz). We do not discuss here all the advantages \cite{Kitching2000,BenAroya2007,Yudin2017,Kobtsev2018,Chuchelov2018,Yudin2023,Tsygankov2025}, which are provided by the PDH technique for MACs over a ``regular'' frequency locking method where $f_m$$\,\ll\,$FWHM. For our current work, it is important just to note that this technique provides additional extrema in the function $\Delta(B_z)$ and their location can be easily controlled by changing $f_m$ \cite{Vishnyakov2025}.

In this Letter, we report on even more brilliant feature of the PDH technique. Using a $0.125$~cm$^3$ vapor cell filled with $^{87}$Rb, we show that there is a ``magic'' combination of $f_m$ and $B_z$, which allows achieving extremely deep suppression of the MFS down to $\Delta$$/$$\Delta_g$$\,\approx\,$$1$$\,\times\,$$10^{-14}$$\delta B$~mG$^{-1}$ over a relatively large interval of the magnetic field variations in the cell ($\approx\,$$6$~mG). Here, $\Delta_g$$\,\approx\,$$6.83$~GHz is the hyperfine splitting frequency in the atomic ground state. Together with other achievements of recent years, the proposed method paves the way for building a new-generation MAC with significantly enhanced long-term frequency stability.

\textit{Qualitative analysis.}---The mitigation of the MFS effect in a MAC by using the PDH technique can be explained by the following qualitative analysis. As a circularly polarized two-frequency light beam is applied to the $^{87}$Rb D$_1$ line in the presence of the longitudinal magnetic field [Fig. \ref{fig:1}(a)], a triple CPT resonance can be observed \cite{Pati2021}. Such a resonance curve can be modeled by three Lorentzian profiles \cite{Tsygankov2021}:

\begin{equation}\label{eq:profile1}
    S(\delta_{\rm R})=\sum_{i=1}^3\frac{A_i\Gamma_i^2}{(\delta_{\rm R}-\delta_i)^2+\Gamma_i^2}\,,
\end{equation}

\noindent where $A_i$, $\Gamma_i$ and $\delta_i$ is the height, width and location of the $i$-th CPT resonance, respectively. In particular, $\delta_2$$\,=\,$$0$ for the central resonance, which comes from the ``0--0'' transition. Other resonances are caused by the magnetically-sensitive two-photon transitions $|F$$=$$1,\,$$m$$=$$-1$$\rangle$$\,\to\,$$|F$$=$$2,\,$$m$$=$$-1$$\rangle$ and $|F$$=$$1,\,$$m$$=$$1$$\rangle$$\,\to\,$$|F$$=$$2,\,$$m$$=$$1$$\rangle$. Eq. (\ref{eq:profile1}) is simulated in Fig. \ref{fig:1}(c).

Each of the resonances in Fig. \ref{fig:1}(c) is connected with a corresponding $\Lambda$-scheme of the atomic energy levels, which are colored in red, green and blue in Fig. \ref{fig:1}(a). For instance, right resonance in Fig. \ref{fig:1}(c) is observed when the Raman frequency detuning ($\delta_{\rm R}$) is equal to double Larmor frequency ($\Omega_{\rm L}$), i.e. $\delta_{\rm R}$$\,=\,$$\nu_{+1}$$-$$\nu_{-1}$$-$$\Delta_g$$\,=\,$$\delta_3$$\,=\,$$2\Omega_{\rm L}$. Here, $\Omega_{\rm L}$$\,=\,$$\gamma$$B_z$ with $\gamma$$\,\approx\,$$700$~Hz/mG being the gyromagnetic ratio for $^{87}$Rb. The optical frequencies $\nu_{+1}$ and $\nu_{-1}$ are $\pm1$ order sidebands, which are presented in the laser radiation spectrum owing to the pump current modulation at a microwave frequency ($\approx\,$$3.4$~GHz). These sidebands are usually used in MACs to pump the atoms into a CPT state \cite{Kitching2018,Skvortsov2020}. The difference in strengths (Rabi frequencies) of the optical transitions in the $\Lambda$-schemes as well as the optical pumping effect lead to the difference in the CPT resonance heights in Fig. \ref{fig:1}(c).

\begin{figure}[!t]
\includegraphics[width=1\linewidth]{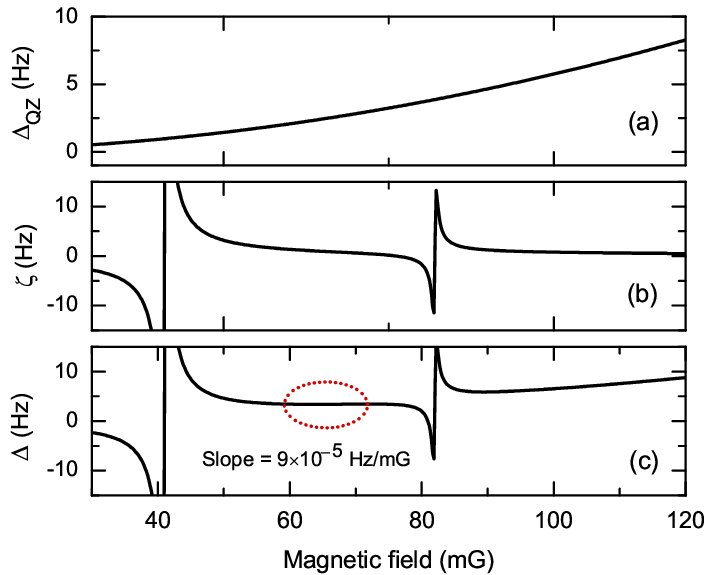}
\caption{\label{fig:2}Theoretical modeling of the magnetic-field-induced frequency shift of the clock (``0--0'') transition in $^{87}$Rb. (a) Quadratic Zeeman shift. (b) Frequency pulling effect from Eq. (\ref{eq:zeta}). (c) Total frequency shift. The flat-shelf peculiarity is outlined with a dotted red oval. Parameters of the modeling are as follows: $\Gamma_{31}$$\,=\,$$\Gamma_{32}$$\,=\,$$0.2$~mG, $C_1$$\,=\,$$6$$C_2$$\,=\,$$30$~mG$\cdot$Hz, $b_2$$\,=\,$$2$$b_1$$\,=\,$$82$~mG.}
\end{figure}

If the PDH technique is employed, then the laser current modulation at $f_m$$\gg$FWHM results in the appearance of additional optical frequency sidebands, such as $\nu_{-1}$$\pm$$f_m$ and $\nu_{+1}$$\pm$$f_m$ [Fig. \ref{fig:1}(b)]. This complex spectrum causes the emergence of many additional CPT resonances simulated in Fig. \ref{fig:1}(d). An interpretation of the multipeak lineshape can be found in Refs. \cite{Tsygankov2024,Vishnyakov2025} and is not the subject of our present Letter. Similarly to (\ref{eq:profile1}), the resonance profile can then be modeled as follows:

\begin{equation}\label{eq:profile2}
    S_{\rm PDH}(\delta_{\rm R})=\sum_{i=1}^3\sum_{j=1}^5\frac{A_{ij}\Gamma_{ij}^2}{(\delta_{\rm R}-\delta_{ij})^2+\Gamma_{ij}^2}\,,
\end{equation}

\noindent Here, $A_{ij}$, $\Gamma_{ij}$ and $\delta_{ij}$ is the height, width and location of the corresponding CPT resonance, respectively.

In a CPT-based clock, the central ``$00$'' resonance is utilized for a microwave frequency stabilization. As mentioned above, the resonance location is affected by the quadratic Zeeman effect. However, another one effect can cause the resonance shift, which also depends on a bias magnetic field strength. Indeed, as shown in Ref. \cite{Tsygankov2021}, the neighboring CPT resonances in Fig. \ref{fig:1}(c), i.e. ``$-10$'' and ``$+10$'', can influence a location of the central ``$00$'' resonance. This effect is known as frequency pulling. Frequency-pulling-like effects manifest in many other areas of laser spectroscopy \cite{Micalizio2019,Yudin20233}.

Let us now analyze a qualitative behavior of the MFS in the case of the PDH technique on the basis of Fig. \ref{fig:1}(d) and Eq. (\ref{eq:profile2}). At some fixed $f_m$, if the magnetic field is decreasing, the ``$+1$$-2$'' resonance starts influencing a location of the clock ``$00$'' resonance. Further decrease in the bias field leads to a similar influence of the ``$+1$$-1$'' resonance on the ``$00$'' resonance location. These actions are conventionally shown by means of pink dashed arrows in Fig. \ref{fig:1}(d). The ``$-1$$+1$'' and ``$-1$$+2$'' resonances (located to the left of the central resonance) also exert influence on the ``$00$'' resonance but with an opposite sign (i.e., they pull the clock transition frequency to the other direction). However, since the ``$+1$$-2$'' and ``$+1$$-1$'' resonances are stronger than their counterparts, ``$-1$$+2$'' and ``$-1$$+1$'', their influence on a location of the ``$00$'' resonance prevails. Therefore, for our qualitative analysis, it is sufficient to consider the influence of only the ``$+1$$-2$'' and ``$+1$$-1$'' resonances. From this qualitative analysis, we can also expect that the frequency pulling should be suppressed in the case of linearly polarized light when the left and right CPT resonances in Fig. \ref{fig:1}(c) have equal heights and linewidths \cite{Mikhailov2010}.

While the function $\Delta_{\rm QZ}(B_z)$ depends on $B_z$ quite smoothly, the contribution $\zeta(B_z)$ from the frequency pulling effect should lead to steep peculiarities in the total function $\Delta (B_z)$, which reads

\begin{equation}\label{eq:sumshift}
    \Delta (B_z) = \Delta_{\rm QZ}(B_z) + \zeta(B_z)\,.
\end{equation}

\noindent A general reasoning makes it clear that $\zeta(B_z)$ has the form of a sum of several dispersion-like functions (this is also confirmed by the experimental data below). Therefore, after interacting with the ``$+1$$-2$'' and ``$+1$$-1$'' resonances during the magnetic field scan, the location of the clock ``$00$'' resonance experiences the frequency pulling effect, which can be qualitatively modeled as follows:

\begin{equation}\label{eq:zeta}
    \zeta(B_z) = \frac{C_1(B_z-b_1)}{\Gamma_{31}^2+(B_z-b_1)^2} + \frac{C_2(B_z-b_2)}{\Gamma_{32}^2+(B_z-b_2)^2}\,.
\end{equation}

\noindent Here, $C_1$ and $C_2$ are the constants, which are proportional to the corresponding heights and widths of the ``$+1$$-2$'' and ``$+1$$-1$'' resonances, i.e. $A_{31}$, $A_{32}$ and $\Gamma_{31}$, $\Gamma_{32}$ in (\ref{eq:profile2}), while shifts $b_1$ and $b_2$ are proportional to the modulation frequency $f_m$.

Fig. \ref{fig:2}(a) shows a smooth parabolic behavior of the function $\Delta_{\rm QZ}(B_z)$. The frequency pulling effect itself is demonstrated in Fig. \ref{fig:2}(b). In the figure, steep jumps at $b_1$$\,=\,$$41$~mG and $b_2$$\,=\,$$82$~mG occur due to the passage of the ``$+1$$-2$'' and ``$+1$$-1$'' resonances through the clock resonance ``$00$''. As shown in Fig. \ref{fig:2}(c), there can be found certain parameters which provide a kind of flat ``shelf'' in the function $\Delta(B_z)$. This shelf does not represent an extremum, however, it is almost a horizontal line in a relatively wide range of the bias field values. The latter means very low sensitivity of the clock transition frequency to small variations in the magnetic field. The magnetic field corresponding to a center of this shelf can serve as a bias field in a MAC.

\begin{figure}[!t]\label{fig:3}
\includegraphics[width=1\linewidth]{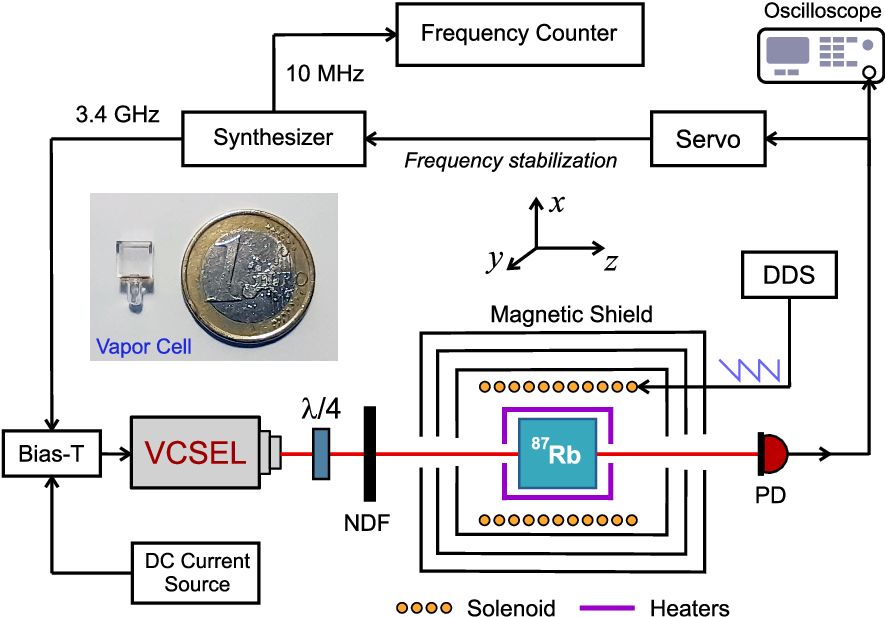}
\caption{Experimental setup: vertical-cavity surface-emitting laser (VCSEL), quarter-wave plate ($\lambda/4$), neutral density filter (NDF), photodetector (PD), direct digital synthesizer (DDS). A cubic $^{87}$Rb vapor cell is shown in the insert.}
\end{figure}

\textit{Experiments.}---As a proof of concept, we have carried out an experiment with a $5$$\times$$5$$\times$$5$~mm$^3$ borosilicate-glass $^{87}$Rb vapor cell filled with argon as a buffer gas ($\approx\,$$100$~Torr). The experimental setup is shown in Fig. \ref{fig:3} with a photo of the cell in the insert.

A commercially available VCSEL with a wavelength of $\approx\,$$794.8$~nm and a spectrum width of $\approx\,$$50$~MHz was employed as the light source. The laser beam diameter ($e^{-2}$) was approximately $1.5$~mm, while the optical power at the cell entrance was maintained at $\approx\,$$17$~$\mu$W using a neutral density filter (NDF). The laser frequency was locked to the center of the absorption profile in the same vapor cell where the CPT resonance was observed. The light had right-circular polarization to induce the $\sigma^+$ transitions in the atom (Fig. \ref{fig:1}).

The vapor cell was enclosed into a three-layer magnetic shield made of permalloy. The residual magnetic field at the center of the shield was compensated with the help of Helmholtz coils installed inside the shield. A homogeneous bias magnetic field ($B_z$) in the cell parallel to the light beam propagation was generated by a solenoid installed inside the shield. The cell temperature was maintained at $\approx\,$$335$~K with an accuracy of about $5$~mK using stabilization electronics and resistive film heaters, which had no noticeable impact on the CPT resonance (the same thermostabilization system was used for high-sensitivity atomic magnetometry \cite{Makarov2025}).

\begin{figure}[!b]
\includegraphics[width=1\linewidth]{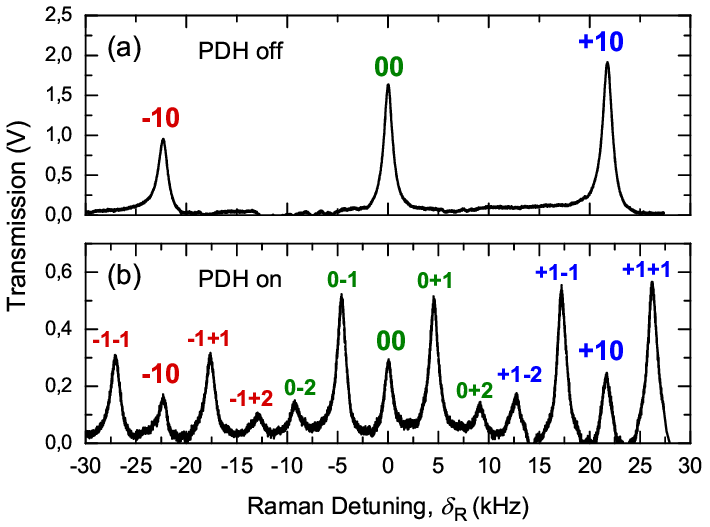}
\caption{\label{fig:4}(a) CPT resonances in a laser beam with right-handed circular polarization in the presence of a bias magnetic field $B_z$$\,=\,$$15$~mG. (b) The same resonances as in (a) but under the fast modulation regime with $f_m$$\,=\,$$4.5$~kHz. The resonance ``00'' at $\delta_{\rm R}$$\,=\,$$0$ corresponds to the clock transition ``0--0''.}
\end{figure}

The laser was driven by a ``Bias-T'' scheme, which allowed mixing the dc pump current with a signal at microwave frequency from the frequency synthesizer to obtain the $\nu_{\pm1}$ sidebands for a CPT resonance excitation. If the PDH stabilization technique is employed, each optical frequency sideband acquired additional low-frequency sidebands, so that the radiation spectrum consisted of the frequencies such as $\nu_{\pm1}$, $\nu_{\pm1}$$\,\pm\,$$f_m$ (the carrier $\nu_0$ and other non-resonant optical sidebands can be omitted from our consideration). For microwave frequency locking to the ``$00$'' resonance, a synchronous modulation-demodulation technique with a modulation frequency $f_m$ was employed. A feedback servo system was established using the ``Red Pitaya'' tool. The 3.4 GHz synthesizer had a 10 MHz output, which was used for frequency shift measurements by means of a frequency counter. A variable voltage with a sawtooth waveform was applied to the solenoid by a direct digital synthesizer (DDS) to study the MFS of the clock transition. Other details on the experimental setup and measurement procedures can be found in Ref. \cite{Vishnyakov2025}.

\begin{figure*}[!t]
\includegraphics[width=0.85\linewidth]{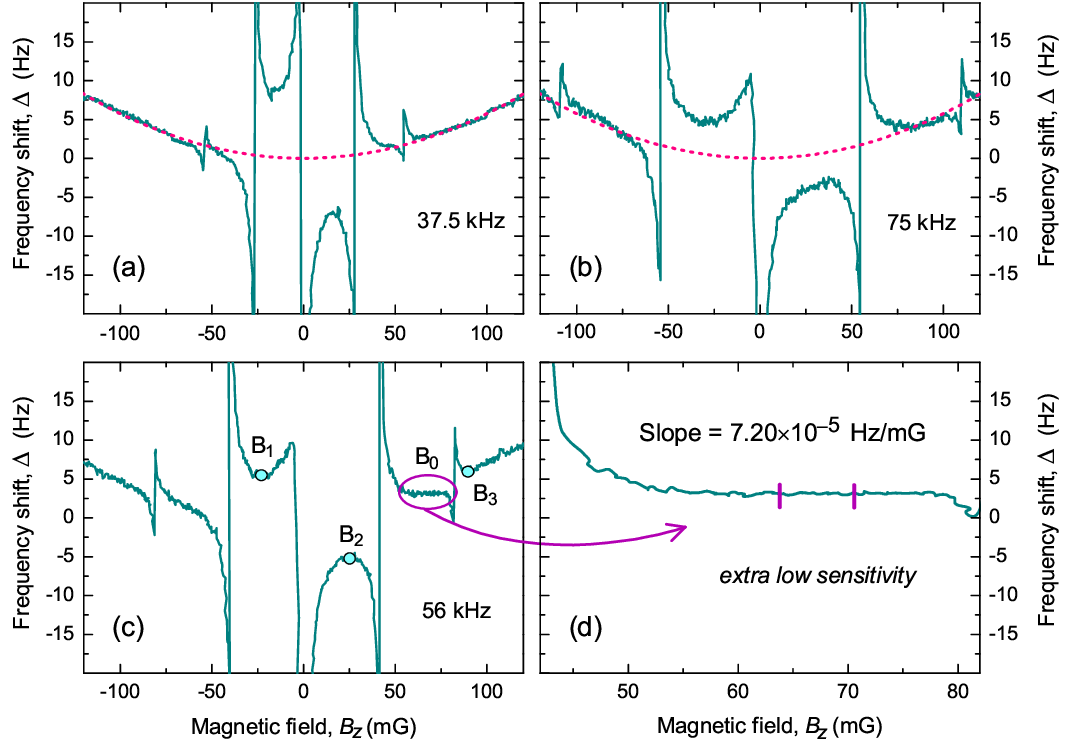}
\caption{\label{fig:5}Frequency shift of the ``0--0'' transition in $^{87}$Rb as the function of a longitudinal magnetic field at $f_m$$\,=\,$$37.5$~kHz (a), $75$~kHz (b), and $56$~kHz (c), (d). The pink dashed line in (a) and (b) represents exclusively the QZ shift according to Eq. (\ref{eq:QZshift}).}
\end{figure*}

In Fig. \ref{fig:4}(a), in the presence of a longitudinal magnetic field $B_z$, the CPT resonance consists of three peaks separated by $\Delta_{\rm R}$$\,=\,$$2\Omega_{\rm L}$ as discussed above. In the PDH regime, as seen in Fig. \ref{fig:4}(b), each of the three resonances splits into five resonances separated by the modulation frequency $f_m$. The location of the left and right groups of five resonances depends linearly on the magnetic field, while the central five resonances experience only the QZ shift. If the magnetic field is properly adjusted, the magnetically-sensitive resonances influence on location of the clock ``$00$'' resonance via the frequency pulling effect. To show this experimentally, we measured the MFS of the clock transition frequency as the function of the magnetic field strength ($B_z$) at three different modulation frequencies ($f_m$). As seen in Fig. \ref{fig:5}, the MFS behavior demonstrates smooth dependence on $B_z$ as well as steep dispersion-like peculiarities owing to the frequency pulling effect. In Fig. \ref{fig:5}(c), some of the extrema are denoted as ${\rm B}_1$, ${\rm B}_2$ and ${\rm B}_3$. Any of these points can serve as a bias magnetic field in a MAC \cite{Vishnyakov2025}. However, in the current study, we are more interested in a specific case when a flat shelf can be formed. This shelf is denoted as ${\rm B}_0$ in Fig. \ref{fig:5}(c). An enlarged image of this section of the function $\Delta(B_z)$ is shown in Fig. \ref{fig:5}(d). The measured slope of the function in the interval from $64$ to $70$~mG is $\approx\,$$7.2$$\times$$10^{-5}$~Hz/mG. This value means that the atomic clock's frequency will have an extremely low sensitivity to small variations in $B_z$, if we set $B_z$$\,\approx\,$$67$~mG and $f_m$$\,\approx\,$$56$~kHz as the operating parameters of the clock.

We have also verified experimentally that the sensitivity of the clock transition frequency to transverse components ($B_x$, $B_y$) of the external magnetic field still obeys the smooth quadratic Zeeman law, similarly to the pink curve in Fig. \ref{fig:5}(a,b). Moreover, in MACs, the shielding factor in the transverse directions is at least one order of magnitude larger than in the longitudinal direction \cite{Hong2020}. Therefore, small perturbations $\delta B_x$ and $\delta B_y$ in the vapor cell cause the fractional frequency $\lesssim\,$$10$$^{-14}$.

\textit{Conclusion.}---We have shown that there is a ``magic'' combination of the modulation frequency and bias longitudinal magnetic field, which provides a sensitivity of the clock transition frequency to small variations in the magnetic field in the vapor cell as low as $\approx\,72$~$\mu$Hz/mG over the $6$ mG interval. It means that the change of direction of the external (Earth's) magnetic field ($\pm$$0.5$~Gauss) will cause the fractional frequency shift ($\Delta$$/$$\Delta_g$) of about $1$$\times$$10$$^{-14}$. Taking into account a shielding factor of $\approx\,$$1000$ in state-of-the-art MACs, the result achieved means that the proposed method allows to reduce the atomic clock's frequency sensitivity to the external magnetic field variations by three orders of magnitude \cite{Datasheets}.

The present work should be considered as a proof-of-principle experiment. Before measuring Allan deviation, a buffer gas mixture in the cell should be optimized. Here, we used rubidium atoms as a benchmark. It is clear that similar results could be obtained with cesium atoms. The revealed effect of the MFS mitigation is the result of interplay between the Zeeman and frequency pulling effects. Therefore, our approach can be used in other types of atomic frequency standards where the Zeeman effect plays a non-negligible role. For instance, the proposed technique could significantly boost the interest to a CPT-based atomic clock scheme, involving ``end resonances'' \cite{Jau2004}. Even more inspiring is the idea of developing a similar method for mitigating the MFS in CPT atomic clocks based on Ramsey-like (pulsed) excitation schemes. Indeed, the light-shift effect in such schemes is significantly suppressed, while the Zeeman effect still poses a significant problem \cite{Mejri2016,Shuker2019}. All these issues, however, are worth separate studies. We believe that, together with other recently proposed methods for suppressing frequency shifts in MACs, the results pave the way for a revolutionary leap in the development of new-generation MACs with significantly enhanced long-term frequency stability in the low $10^{-13}$ range.

%\begin{acknowledgments}
%\end{acknowledgments}

\textit{Acknowledgments.}---We thank A.V.~Taichenachev for carefully reading the manuscript and helpful remarks. The work was supported by the Russian Science Foundation (Grant No. 23-12-00195).

%\bibliography{apssamp}
\bibliography{Manuscript}
% Produces the bibliography via BibTeX.

\end{document}